\documentclass[a4paper]{jpconf}

\usepackage{graphicx,slashed}
\usepackage{epsf}
\usepackage{amsfonts}
\usepackage{amsmath}
\usepackage{bm}
\newcommand{\be}{\begin{equation}}\newcommand{\ee}{\end{equation}}
\newcommand{\bea}{\begin{eqnarray}}\newcommand{\eea}{\end{eqnarray}}
\newcommand{\brr}{\begin{array}}\newcommand{\err}{\end{array}}
\newcommand{\bit}{\begin{itemize}}\newcommand{\eit}{\end{itemize}}
\newcommand{\ben}{\begin{enumerate}}\newcommand{\een}{\end{enumerate}}

\newcommand{\bbm}{\begin{bmatrix}}\newcommand{\ebm}{\end{bmatrix}}
\newcommand{\ba}{\begin{array}}
\newcommand{\ea}{\end{array}}
\newcommand{\G}{\textbf}

\newtheorem{mydef}{Definition}
\newtheorem{Lemma}{Lemma}
\newtheorem{theorem}{Theorem}
\newcommand{\bd}{\begin{mydef}} \newcommand{\ed}{\end{mydef}}
\newcommand{\bthe}{\begin{theorem}} \newcommand{\ethe}{\end{theorem}}
\newcommand{\ble}{\begin{Lemma}} \newcommand{\ele}{\end{Lemma}}

\newcommand{\dr}{\mathrm{d}}

\def\ha{\frac{1}{2}}

\def\intx{\int \!\!\mathrm{d}^3 x}

\def\lab{\label}\def\lan{\langle}
\def\lf{\left}

\def\non{\nonumber}\def\pa{\partial}\def\ran{\rangle}

\def\ri{\right}
\def\al{\alpha}\def\bt{\beta}\def\ga{\gamma}
\def\de{\delta}\def\De{\Delta}

\def\la{\lambda}\def\La{\Lambda}\def\si{\sigma}
\def\om{\omega}

\def\1{{_{1}}}\def\2{{_{2}}}
\newcommand{\ide}{1\hspace{-1mm}{\rm I}}

\def\noHe0{:\;\!\!\;\!\!:H_e(0):\;\!\!\;\!\!:}
\def\noHm0{:\;\!\!\;\!\!:H_\mu(0):\;\!\!\;\!\!:}

\def\lab{\label}
\def\lan{\langle}
\def\lf{\left}

\def\non{\nonumber}
\def\pa{\partial}\def\ran{\rangle}

\def\ri{\right}

\def\al{\alpha}\def\bt{\beta}\def\ga{\gamma}
\def\de{\delta}\def\De{\Delta}

\def\la{\lambda}
\def\La{\Lambda}\def\si{\sigma}
\def\om{\omega}

\def\1{{_{1}}}\def\2{{_{2}}}

\begin{document}

\title{Flavor neutrinos as unstable particles: an interaction picture view}

\author{M Blasone $^{1,2}$, F Giacosa $^{3,4}$. L Smaldone $^{1,2}$ and G Torrieri $^{5}$}

\address{$^1$ Physics Department ``E.R. Caianiello'', University of Salerno,\\
  Via Giovanni Paolo II, 132, 84084 Fisciano (Salerno), Italy}
\address{$^2$ INFN Sezione di Napoli, Gruppo collegato di Salerno, Italy}
\address{$^3$ Institute of Physics, Jan-Kochanowski University, ul. Uniwersytecka 7, 25-406 Kielce, Poland}
\address{$^4$ Institute for Theoretical Physics, J. W. Goethe University, Max-von-Laue-Stra\ss e 1,
60438 Frankfurt, Germany}
\address{$^5$ Instituto de Fisica Gleb Wataghin - UNICAMP, 13083-859, Campinas SP, Brazil}

\ead{blasone@sa.infn.it}
\ead{francesco.giacosa@gmail.com}
\ead{lsmaldone@unisa.it}
\ead{torrieri@unicamp.br}

\begin{abstract}
This paper reviews the similarities in the behavior of unstable particles and oscillating neutrinos using perturbation theory within the interaction picture of quantum field theory. We begin by examining how decaying systems are studied in the interaction picture and then demonstrate how similar calculations can be performed to determine the transition probabilities for flavor oscillations. Notably, the expressions for neutrino oscillations and particle decays are identical in the short-time range. Furthermore, we show that the flavor oscillation formula derived through this method matches, within the adopted approximation, the one obtained using the flavor Fock space approach.
\end{abstract}

\section{Introduction}
Some of our previous work highlighted the connection between unstable particles and oscillating neutrinos \cite{Blasone2019,Blasone:2020qbo,Blasone2020,Luciano:2023yzt}. The key ingredient in both cases is the time-energy uncertainty relation. This uncertainty plays a crucial role in both the decay of unstable particles and the way neutrinos change flavors as they travel. In the case of unstable particles, the uncertainty in time is directly related to their lifetime \cite{Bhattacharyya_1983}. For neutrinos, it is connected to their oscillation length.
It is interesting to note that the so-called \emph{external wave-packet} approach \cite{Giunti:1993se}, which treats neutrinos as internal lines of macroscopic Feynman diagrams, also has its roots in the study of unstable particles \cite{Jacob1961}.

Building on these ideas, our recent work \cite{Blasone:2023brf} explored how perturbation theory in the interaction picture within quantum field theory (QFT) can be used to describe flavor oscillations. This approach offers a fresh perspective by treating the part of the theory responsible for flavor mixing as an interaction between different neutrino flavors. We performed calculations for both bosons and fermions in various spacetime dimensions (0+1 and 3+1). Remarkably, the flavor transition and survival probabilities we obtained coincide with that computed in the \emph{flavor Fock space} approach to neutrino mixing and oscillations in QFT \cite{Blasone:1995zc,BHV99,Blasone:2001qa,Smaldone:2021mii}.

In the present work, which can be viewed as a sort of second part of Ref. \cite{Blasone:2020qbo}, we review some main results of Ref. \cite{Blasone:2023brf}. Before doing it, we review how the same machinery has been applied to study decaying systems \cite{PhysRevLett.71.2687,facchi1999regola}, deriving the exponential decay law for large times and the quantum Zeno effect for small times.

The paper is organized as follows. In Section \ref{secun}, the treatment of unstable systems within perturbation theory is briefly reviewed. Then, in Section \ref{r2}, the interaction picture view of flavor oscillations is also briefly presented. In particular we report the example of flavor transitions between two-flavors for Dirac fields in (3+1)D. Finally, in Section \ref{conc} conclusions are presented.

\section{Decay of unstable particles in the interaction picture} \label{secun}
Suppose we want to describe a decay process. In the interaction (Dirac) picture, the Hamiltonian can be decomposed as
\be
H \ = \ H_0 \ + \ H_{int} \, .
\ee
If the decaying particle state at some time $t=0$ is described by the state $|a\ran$, which is an eigenstate of the free part of the Hamiltonian
\be
H_0 |a\ran \ = \ E_a |a \ran \, , 
\ee
the survival amplitude can be written as
\be \label{ampl}
\mathcal{A} \ = \ e^{-i E_a t} \, \lan a|U(t,0)|a\ran \, , 
\ee
where the time evolution operator can be expressed by the Dyson series
\be \label{dyfor1}
U(t,0) \ = \ \mathcal{T}\exp  \lf[-i \int^{t}_{0} \!\! \dr t \, H_{int}(x) \ri] \, .
\ee
Here $\mathcal{T}$ is the chronological product. Suppose that $H_{int}$ contains a coupling constant which allows us to employ perturbative computations. Then we can expand the previous expression up to the second order
\be \label{dyfor2}
U(t,0) 
\ = \ 1-i\int_{0}^{t}\!\!\dr t_{1} \,H_{int}(t_{1})-
\int_{0}^{t}\!\!\dr t_{1} \,
\int_{0}^{t_1}\dr t_{2} \,[H_{int}(t_{1}) \,  H_{int}(t_{2})+...
\ee
Splitting the Hamiltonian so that the interaction term has no diagonal contributions
\be
\lan a|H_{int}|a\ran \ = \ 0 \, , 
\ee
the first order contribution to the amplitude \eqref{ampl} erases. Then
\be
\mathcal{A} \ = \ e^{-i E_a t} \ - \ e^{-i E_a t}
\int_{0}^{t}\!\!\dr t_{1} \,
\int_{0}^{t_1}\dr t_{2} \,\lan a|H_{int}(t_{1}) \,  H_{int}(t_{2})|a\ran \, .
\ee
In order to further develop the second order piece, we can insert the resolution of the identity in terms of final states
\be
\ide \ = \  |f\ran \lan f| \,  , \quad H_0|f\ran \ = \ E_f |f\ran \, , 
\ee
and perform the time integrals, getting \cite{facchi1999regola}
\be
\mathcal{A} =  e^{-i E_a t}\lf\{1  -  \, \sum_{f \neq a} \, |\lan f|H_{int}|a\ran|^2 \, \lf[\frac{2 \sin^2\lf(\frac{E_f-E_a}{2}t\ri)}{(E_f-E_a)^2}+i \lf(\frac{1}{E_f-E_a}+\frac{\sin\lf[\lf(E_f-E_a\ri)t\ri]}{(E_f-E_a)^2}\ri)\ri]\ri\} \, .
\ee
The survival probability is thus obtained taking the square $\mathcal{P} \ = \ |\mathcal{A}|^2$
\be
\mathcal{P} =  1  -  \, \sum_{f \neq a} \,  |\lan f|H_{int}|a\ran|^2 \,  \frac{\sin^2\lf(\frac{E_f-E_a}{2}t\ri)}{\lf(\frac{E_f-E_a}{2}\ri)^2} \, .
\ee
For an unstable system the final energy spectrum is continuous, so that the previous expression has to be written as
\be \label{pe}
\mathcal{P} =  1  -  \int \!\! \dr E \, \rho(E) \, |\lan E|H_{int}|a\ran|^2 \, \frac{\sin^2\lf(\frac{E-E_a}{2}t\ri)}{\lf(\frac{E-E_a}{2}\ri)^2} \, ,
\ee
where $\rho(E) \equiv \sum_f \de (E-E_f)$ is the final density of states. When $t \to \infty$, one can use that
$
\lim_{t \to \infty} \, \frac{\sin^2\lf(\frac{E-E_a}{2}t\ri)}{\lf(\frac{E-E_a}{2}\ri)^2} \ = \ t \pi \de(E-E_a) 
$
getting
\be
\mathcal{P} =  1  -  \ga \, t \ \approx \ e^{-\ga t} \, ,  \, ,
\ee
where $\ga$ is given by the \emph{Fermi golden rule}
\be
\ga \ = \  2 \, \pi \, \rho(E_a) |\lan E_a|H_{int}|a\ran|^2 \, .
\ee

Actually the previous reasoning has some important limitations. Firstly, to ensure the vacuum stability, the energy must be bounded from below. Moreover, we took the $t \to \infty$ limit and then the short-time behavior could be drastically different from the exponential decay law  \cite{PhysRevLett.71.2687,facchi1999regola,Giacosa:2010br,Giacosa:2011xa,Giacosa:2018dzm,Giacosa:2021hgl} . In fact, taking the lower energy equal to zero and expanding Eq.\eqref{pe} around $t=0$ 
\be \label{pest}
\mathcal{P} =  1  - t^2  \int^\La_0 \!\! \dr E \, \rho(E) \, |\lan E|H_{int}|a\ran|^2 \ = \ 1-\frac{t^2}{ \tau_Z^2} \,  \, ,
\ee
where $\La$ is an ultraviolet cut-off. $\tau_Z$ is the \emph{Zeno time} and the Eq.\eqref{pest} expresses the \emph{quantum Zeno effect}: because the decay probability is negligible at short-time, a quantum system frequently observed soon after the production time never decays. 

As an explicit example one can consider a superrenormalizable theory with
\be
H_{int} \ = \ \intx \, \mathcal{H}_{int}(x) \, , \quad \mathcal{H}_{int} \ = \ \la \, \phi_1(x) \, \phi_2(x) \, \psi(x) \, , 
\ee
which describes the decay $\psi \to \phi_1 \, \phi_2$. If the initial $\psi$ state has an energy at rest $E_a=M$, one can compute \cite{PhysRevLett.71.2687,facchi1999regola}
\be
\frac{1}{\tau^2_Z}\ = \ \frac{\la^2}{16 \pi^2 M} \lf(\La-M\ri) \, . 
\ee
Note that this expression diverges when the cut-off is removed ($\La \to \infty$), thus $\tau_{Z}$ vanishes. This is due to due to the local nature of the interaction. However, it does not mean that the decay is exponential: as shown in Ref. \cite{Giacosa:2010br}, the first derivative of the survival probability vanishes, implying that the Zeno mechanism is possible, even if the Zeno-time tends to zero.  The main difference is that a full resummation of one-loop diagrams is necessary to obtain  a viable spectral function \cite{Giacosa:2007bn}, a feature that cannot be  obtained if a perturbative calculation up to second order -as the one outlined above- is implemented.  
Moreover, in certain physically interesting cases, the parameter $\La$ is finite \cite{facchi1999regola}.
\section{Neutrino oscillations in the interaction picture} \label{r2}
Let us consider the Hamiltonian density
\bea \lab{neutr}
\mathcal{H} \  =  \ \mathcal{H}_{0} \, + \, \mathcal{H}_{int} \, ,
\label{Linteract}
\eea
with
\bea
\mathcal{H}_{0} & = & \sum_{\si=e,\mu} \lf[\overline{\nu}_\si \lf(i \slashed{\pa}-m_\si\ri) \nu_\si \ri] \\[2mm]
 \mathcal{H}_{int} & = & m_{e \mu} \lf(\overline{\nu}_e \nu_\mu+\overline{\nu}_\mu \nu_e\ri) \, , 
\eea
which can be used to describe neutrino oscillations. In Ref. \cite{Blasone:2023brf} we studied the time evolution operator 
\be \label{dyfor}
U(t_i,t_f) \ = \ \mathcal{T}\exp  \lf[-i \int^{t_f}_{t_i} \!\! \dr^4 x \, :\mathcal{H}_{int}(x): \ri] \, ,
\ee
where $::$ indicates normal ordering. As in the previous section we will stop at the second order in $m_{e \mu}$.

\begin{figure}[h]
        \centering       \includegraphics[scale=0.6]{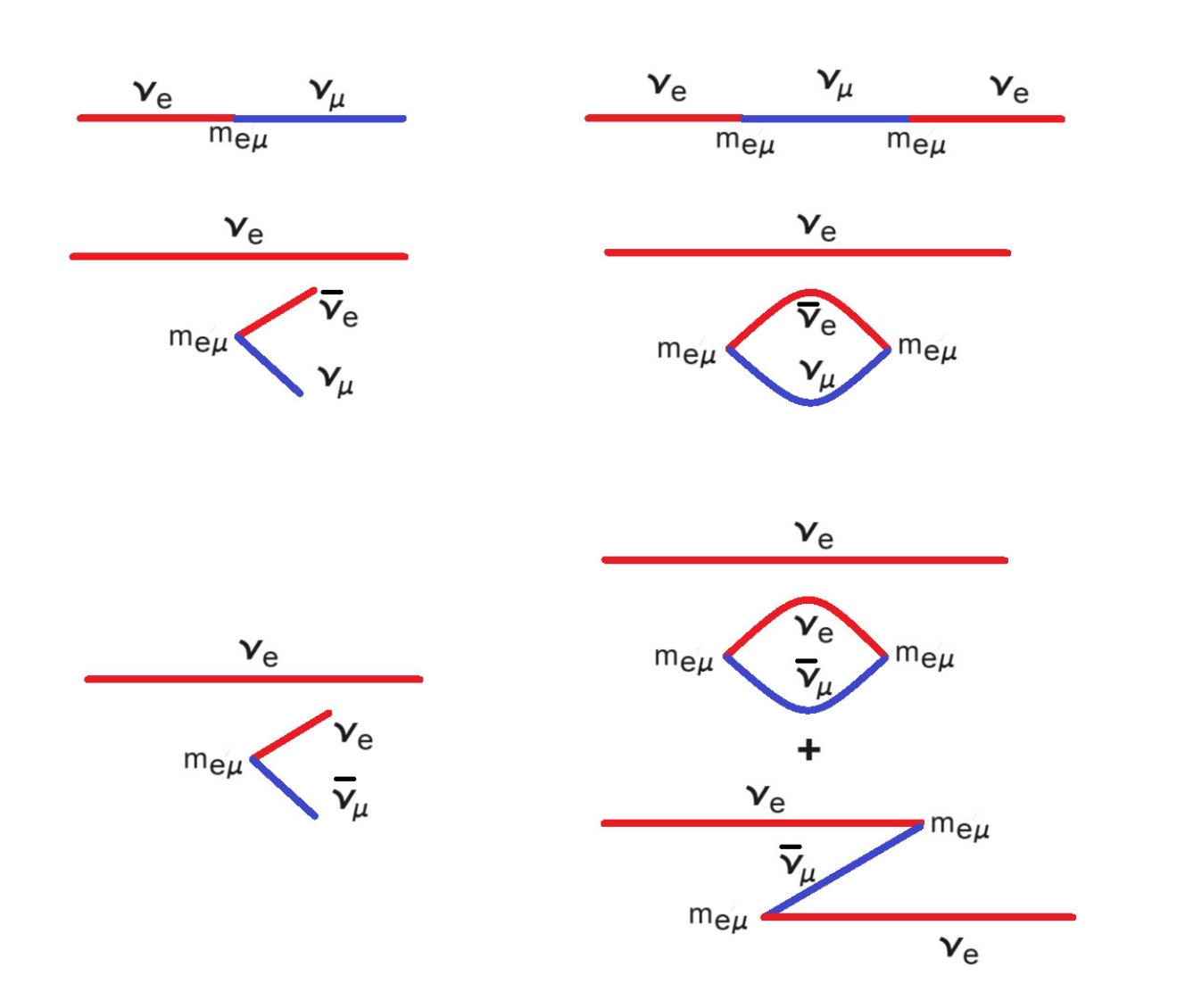}\\
        \caption{Schematic representation of the different contributions. Left column: the transition $\nu_e$ into $\nu_{\mu}$ at first order; right: the corresponding second order diagrams, whose imaginary part reduces to the contribution on the left. In particular: (i) The first diagram on the left is the direct transition $\nu_e \rightarrow \nu_{\mu}$, see Eq. (\ref{wcon}). This is the usual oscillation involving the difference of the energies.  (ii) The second diagram on the left corresponds to Eq. \eqref{second}, which vanishes since it needs to be subtracted. Namely, as the r.h.s. shows, this is a disconnected vacuum diagram.  (iii) The third diagram on the left corresponds to Eq.(\ref{thirda}), which consists of a piece that needs to be subtracted, and a piece that survives, see the two survival diagrams on the r.h.s. as well as Eq. (\ref{thirdb}), in which the sum of energies enters.  }
        \label{diagrams}
   \end{figure}

In the interaction picture $\nu_\si$ ($\si=e,\mu$) can be expanded as
\begin{eqnarray}
\nu_{\si}(x) = \frac{1}{\sqrt{V}} \sum_{\G k,r}\,  \left[ u_{{\bf k},\si}^{r}(t) \, \alpha_{{\bf k},\si}^{r} + v_{-{\bf k},\si}^{r}(t) \, \bt_{-{\bf k},\si}^{r\dag}   \right]  e^{i{\bf k}\cdot {\bf x}}  \, ,
\label{fieldex}
\end{eqnarray}
with $u^r_{{\bf k},\si}(t) \,= \, e^{- i \om_{\G k,\si} t}\, u^r_{{\bf k},\si}\;$,
$\;v^r_{{\bf k},\si}(t) \,= \, e^{ i \om_{\G k,\si} t}\, v^r_{{\bf k},\si}$,
 $\om_{\G k,\si}=\sqrt{|\G k|^2 + m_\si^2}$. The perturbative vacuum is defined by
\be \label{vacm}
\al^r_{\G k, \si}|0 \rangle = 0 = \beta _{{\bf k},\si}^{r} |0 \rangle \  .
\ee
Here the ladder operators satisfy the anticommutation relations
\be  \label{CAR2} \{\al ^r_{{\bf k},\rho}, \al ^{s\dag }_{{\bf q},\si}\} = \de_{\G k \G q}\de _{rs}\de _{\rho \si}  \quad \, , \quad \{\bt^r_{{\bf k},\rho},
\bt^{s\dag }_{{\bf q},\si}\} =
\de_{\G k \G q} \de _{rs}\de _{\rho \si}. 
\ee
Then, one can compute $H_{int}$ as
\bea
H_{int}(t)& = &  m_{e \mu}  \sum_{s,s'=1,2}\sum_{\G p}
 \Big[\bt^s_{\G p,\mu}\bt^{s\dag}_{\G p,e} \de_{s s'} W^*_\G p(t)+\al^{r\dag}_{\G p,\mu} \al^r_{\G p,e} \de_{s s'} W_\G p(t) \non \\[2mm]
& + & \bt^s_{-\G p,\mu}\al^{s'}_{e,\G p} \lf(Y^{s s'}_\G p(t)\ri)^*+\al^{s\dag}_{\G p,\mu}\bt^{s'\dag}_{-\G p,e} Y^{s s'}_\G p(t)\, + \, e \leftrightarrow \mu \Big] \, ,
\eea
where we defined
\bea
W_\G p(t) & = & \overline{u}^s_{\G p,\mu} u^s_{\G p,e} e^{i\left( \omega_{\G k,\mu}-\omega_{\G k,e}\ri)t} \ = \ W_\G p \,  e^{i\left( \omega_{\G p,\mu}-\omega_{\G p,e}\ri)t}  \\[2mm]
Y^{s s'}_\G p(t) & = &  \, \overline{u}^{s}_{\G p,\mu} v^{s'}_{-\G p,e} e^{i\left( \omega_{\G k,\mu}+\omega_{\G k,e}\ri)t} \ = \ Y^{s s'}_\G p e^{i\left( \omega_{\G p,\mu}+\omega_{\G p,e}\ri)t}
\eea
Their explicit expression is
\bea
W_\G p & = & \sqrt{\frac{\lf(\omega_{\G p,e}+m_{e}\ri)\lf(\omega_{\G p,\mu}+m_{\mu}\ri)}{4\omega_{\G p,e} \omega_{\G p,\mu}}}
\left(1-\frac{|\G p|^{2}}{(\omega_{\G p,e}+m_{e})(\omega_{\G p,\mu}+m_{\mu})}\right)  \, , \\[2mm]
Y^{22}_{\G p} & = & -Y^{11}_{\G p} \ = \ \frac{p_3}{\sqrt{4 \om_{\G p,e}\om_{\G p,\mu}}}
\lf(\sqrt{\frac{\om_{\G p,\mu}+m_\mu}{\om_{\G p,e}+m_e}}+\sqrt{\frac{\om_{\G p,e}+m_e}{\om_{\G p,\mu}+m_\mu}}\ri) \, , \\[2mm]
Y^{12}_{\G p} & = & \lf(Y^{21}_{\G p} \ri)^* \ = \ -\frac{p_1-i p_2}{\sqrt{4 \om_{\G p,e}\om_{\G p,\mu}}}
\lf(\sqrt{\frac{\om_{\G p,\mu}+m_\mu}{\om_{\G p,e}+m_e}}+\sqrt{\frac{\om_{\G p,e}+m_e}{\om_{\G p,\mu}+m_\mu}}\ri) \, .
\eea
A non-trivial process which contributes to flavor oscillation amplitude is (see Fig. \ref{diagrams}, first row):
\be
|\nu^r_{\G p,e}\ran \ \rightarrow \ |\nu^s_{\G k,\mu}\ran \, , \qquad |\nu^r_{\G p,\si}\ran \equiv \al^{r\dag}_{\G p,\si}|0\ran \, .
\ee
The amplitude of this process reads, at the first order, as
\bea \non
{}\hspace{-3mm} \mathcal{A}^{rs}_{e \to \mu}(\G p,\G k, ;t_i,t_f) \ &\approx & \ - i m_{e \mu}\de_{r s} \de_{\G k,\G p} W_{\G p}\, \ \int^{t_f}_{t_i} \!\! \dr t \, e^{i\lf(\om_{\G k,\mu}-\om_{\G p,e}\ri)t} \non \\[2mm]   
& = & m_{e \mu} \,  \de_{r s} \de_{\G k,\G p}  \, \,  \lf(e^{i\lf(\om_{\G p,\mu}-\om_{\G p,e}\ri)t_f}-e^{i \lf(\om_{\G p,\mu}-\om_{\G p,e}\ri)t_i}\ri)  \frac{W_{\G p}}{\om_{\G k,e}-\om_{\G k,\mu}} \non \\[2mm]
 & = &  \de_{r s} \de_{\G k,\G p}  \,  \tilde{\mathcal{A}}_{e \to \mu}(\G k; t_i,t_f) \, , 
\eea
where
\be
\tilde{\mathcal{A}}_{e \to \mu}(\G p; t_i,t_f) \ = \ \frac{m_{e \mu} \, W_{\G p}}{\om_{\G p,e}-\om_{\G p,\mu}} \, \lf(e^{i\lf(\om_{\G p,\mu}-\om_{\G p,e}\ri)t_f}-e^{i \lf(\om_{\G p,\mu}-\om_{\G p,e}\ri)t_i}\ri) \, . 
\ee
To compute the oscillation probability we must sum over the final density of states
\bea \non
\mathcal{P}_{e \to \mu}(\G p;\De t) & = & \sum_{\G k,s} |\mathcal{A}^{rs}_{e \to \mu}(\G p, \G k;t_i,t_f)|^2 \ = \ |\tilde{\mathcal{A}}_{e \to \mu}(\G p, t_i,t_f)|^2 \\[2mm]
& = &  W_{\G p}^2 \, \frac{2 m^2_{e \mu} }{\lf(\om_{\G p,e}-\om_{\G p,\mu}\ri)^2}  \lf[1-\cos\lf[\lf(\om_{\G p,\mu}-\om_{\G p,e}\ri)\De t\ri] \ri]  \, , \qquad \ \De t \equiv t_f-t_i \, . \label{wcon}
\eea
\begin{figure}[h]
        \centering       \includegraphics[scale=0.55]{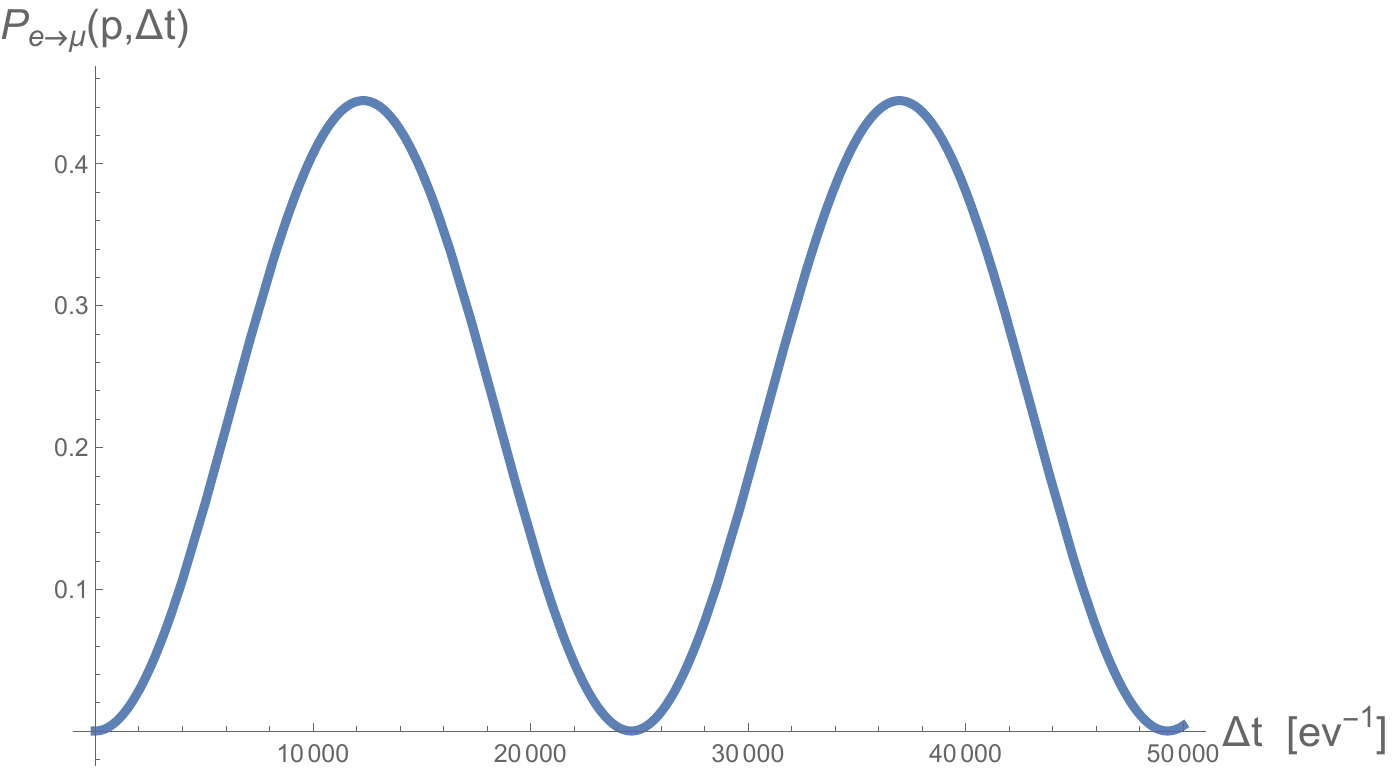}\\
        \caption{Plot of the function $\mathcal{P}_{e \to \mu}(\G p;\De t)$ of Eq. \eqref{wcon} as function of the time interval $\De t$ for  the neutrino masses $m_{e} =0.07 $ eV, $m_{\mu} = 0.1 $ eV, for the mixing parameter $m_{e \mu} = 0.01$ eV, and for the modulus of the three-momentum $p = 10$ eV.  }
        \label{pemu}
   \end{figure}
We refer to Fig. \ref{pemu} for a numerical example of the probability of this process.
   
   \begin{figure}[h]
        \centering       \includegraphics[scale=0.55]{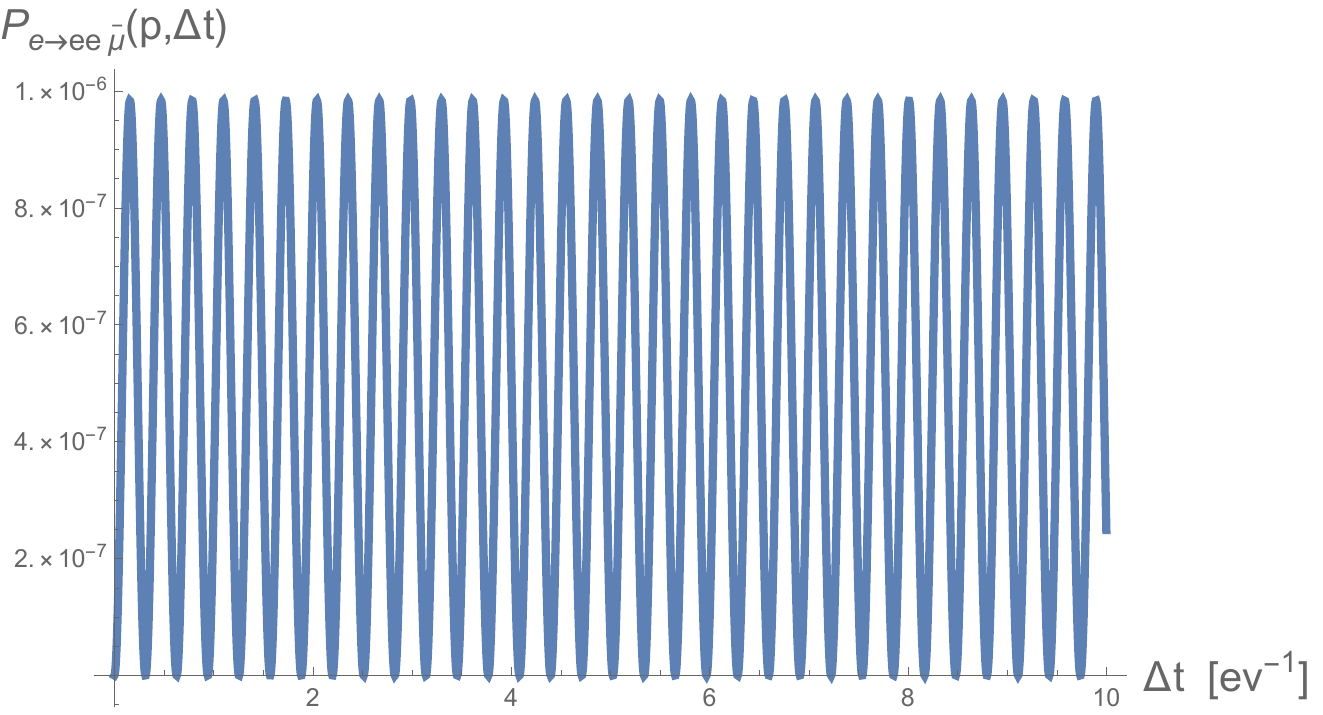}\\
        \caption{Plot of the function $\mathcal{P}_{e \to e e \overline{\mu}}(\G p;\De t) $ of Eq. \eqref{thirdb} as function of the time interval $\De t$ for  the same parameters displaced in the caption of Fig. \ref{pemu}. The frequency of oscillations is much larger but the intensity much smaller than those of  Fig. \ref{pemu}. }
        \label{peeantimu}
   \end{figure}

A second non-trivial process which should be considered for the transition amplitude is (Fig. \ref{diagrams}, second row)
\be
|\nu^r_{\G p,e}\ran \ \rightarrow \ |\nu^{s_1}_{\G k_1,e}\ran |\nu^{s_2}_{\G k_2,\mu}\ran |\overline{\nu}^{s_3}_{\G k_3,e}\ran \, .
\ee
One finds that
\bea \non
&& \mathcal{A}^{r s_1 s_2 s_3}_{e \to e\overline{e} \mu}(\G p,\G k_1,\G k_2, \G k_3;t_i,t_f)  \approx
 -i \, m_{e \mu} \, \, Y_{\G k_2}^{s_3 s_2} \, \de_{\G k_1, \G p}  \de_{\G k_2, - \G k_3}\, \de_{r s_1}   \int^{t_f}_{t_i} \!\! \dr t \, e^{-i\lf(\om_{\G k_2,\mu}+\om_{\G k_2,e}\ri)t}   \\[2mm] \non
&& = \  -m_{e \mu}  \, \de_{r s_1} \,  \de_{\G k_1, \G p}  \de_{\G k_2, - \G k_3} \,  \lf(e^{-i\lf(\om_{\G k_2,\mu}+\om_{\G k_2,e}\ri)t_f}-e^{-i \lf(\om_{\G k_2,\mu}+\om_{\G k_2,e}\ri)t_i}\ri)  \frac{Y_{\G k_2}^{s_2 s_3}}{\om_{\G k_2,e}+\om_{\G k_3,\mu}} \\
&& =  \ \de_{\G k_1, \G p}  \de_{\G k_2, - \G k_3} \,  \de_{r s_1} \, \tilde{\mathcal{A}}^{s_2 s_3}_{e \to e\overline{\mu}\mu}(\G k_2;t_i,t_f) \, , 
\eea
where
\be
\tilde{\mathcal{A}}^{s_2 s_3}_{e \to e\overline{e} \mu }(\G k;t_i,t_f) \ = \ -\frac{m_{e \mu} \, Y^{s_2 s_3}_{\G k}}{\om_{\G k,e}+\om_{\G k,\mu}} \, \lf(e^{-i\lf(\om_{\G k,\mu}+\om_{\G k,e}\ri)t_f}-e^{-i \lf(\om_{\G k,\mu}+\om_{\G k,e}\ri)t_i}\ri) \, . 
\ee
We thus get
\be
\mathcal{P}_{e \to e\overline{e} \mu}(\G p;\De t)  \ = \ \sum_{\G k_1,\G k_2,\G k_3} \sum_{s_1,s_2,s_3} |\mathcal{A}^{r s_1 s_2 s_3}_{e \to e\overline{e} \mu}(\G p,\G k_1,\G k_2, \G k_3;t_i,t_f) |^2  \ = \ \sum_{\G k}\sum_{s_2,s_3}|\tilde{\mathcal{A}}^{s_2 s_3}_{e \to e\overline{e} \mu }(\G k;t_i,t_f)|^2\, .
\ee
In the large-$V$ limit we obtain
\be
\mathcal{P}_{e \to e\overline{e} \mu}(\G p;\De t)  \ = \ V \sum_{s_2,s_3} \, \int \!\! \frac{\dr^3 \G k}{(2 \pi)^3} \, \frac{\lf(Y^{s_2 s_3}_\G k\ri)^2 }{\lf(\om_{\G k,e}+\om_{\G k,\mu}\ri)^2}  \sin^2\lf(\frac{\lf(\om_{\G k,\mu}+\om_{\G k,e}\ri)\De t}{2}\ri) \, . 
\label{second}
\ee
This infrared divergent contribution only redefines the vacuum energy and must be subtracted from the final result.

\begin{figure}[h]
        \centering       \includegraphics[scale=0.65]{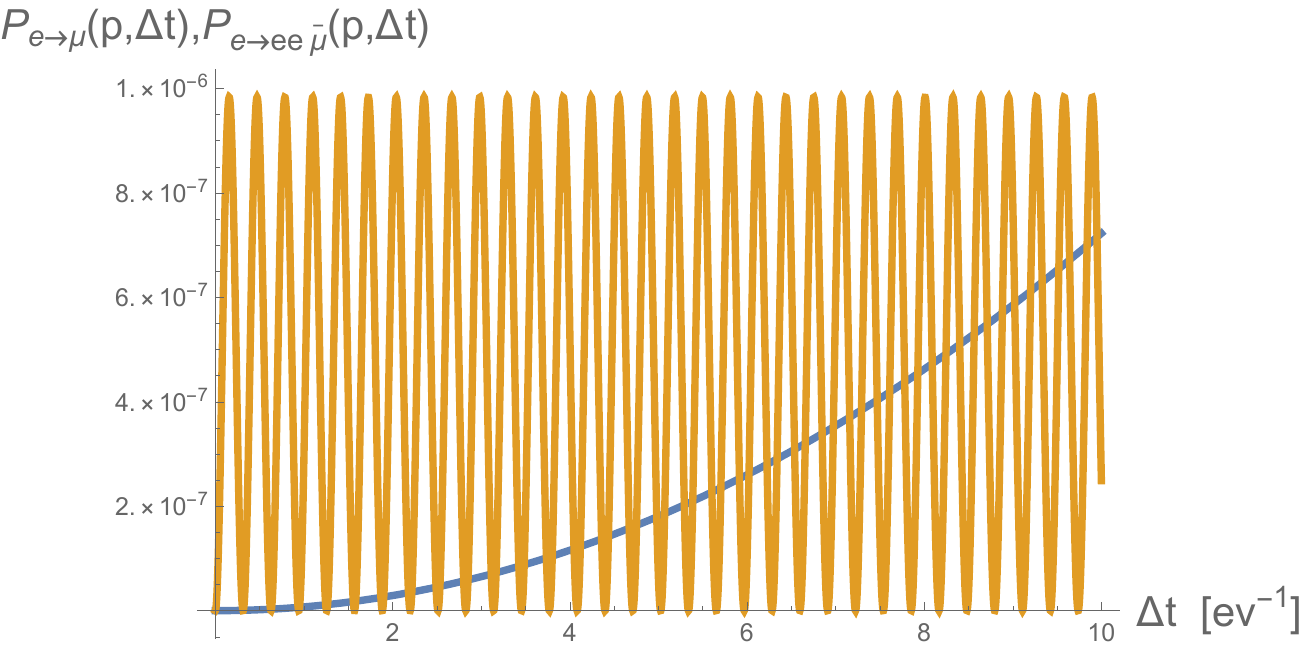}\\
        \caption{Comparison between the functions $\mathcal{P}_{e \to \mu}(\G p;\De t)$ of Eq. \eqref{wcon} (blue, slowly growing) and $\mathcal{P}_{e \to e e \overline{\mu}}(\G p;\De t) $ of Eq. \eqref{thirdb} (yellow, fast oscillating) for a short initial time interval.}
        \label{comparison}
   \end{figure}

The last process to consider is (Fig. \ref{diagrams}, third row)
\be
|\nu^r_{\G p,e}\ran \ \rightarrow \ |\nu^{s_1}_{\G k_1,e}\ran |\nu^{s_2}_{\G k_2,e}\ran |\overline{\nu}^{s_3}_{\G k_3,\mu}\ran \, , \qquad \G k_1 \neq \G k_2 \ \lor \ s_1 \neq s_2 \, ,
\ee
whose amplitude reads
\bea \mathcal{A}^{r s_1 s_2 s_3}_{e \to e e \overline{\mu}}(\G p,\G k_1,\G k_2, \G k_3;t_i,t_f)
& = & \de_{\G k_1, \G p}  \de_{\G k_2, - \G k_3} \,  \de_{r s_1} \, \tilde{\mathcal{A}}^{s_2 s_3}_{e \to e e\overline{\mu}}(\G k_2;t_i,t_f) \non \\[2mm]
& - & \de_{\G k_2, \G p}  \de_{\G k_1, - \G k_3} \,  \de_{r s_2} \, \tilde{\mathcal{A}}^{s_1 s_3}_{e \to e e\overline{\mu}}(\G k_1;t_i,t_f) \, . 
\eea
Here $\tilde{\mathcal{A}}^{s_2 s_3}_{e \to e e \overline{\mu} }(\G k;t_i,t_f)=\tilde{\mathcal{A}}^{s_2 s_3}_{e \to e\overline{e} \mu }(\G k;t_i,t_f)$. 
Then, the probability of such process is given by the expression
\bea
&& \mathcal{P}_{e \to e e \overline{\mu}}(\G p;\De t)  \ = \ \ha \sum_{\G k_1,\G k_2,\G k_3} \sum_{s_1,s_2,s_3} |\mathcal{A}^{r s_1 s_2 s_3}_{e \to e e \overline{\mu}}(\G p,\G k_1,\G k_2, \G k_3;t_i,t_f) |^2 \non \\[2mm]
&& =  \sum_{\G k,s_2,s_3} |\tilde{\mathcal{A}}^{ s_2 s_3}_{e \to e e\overline{\mu}}(\G k;t_i,t_f)|^2-\sum_{s_3} |\tilde{\mathcal{A}}^{r s_3}_{e \to e e\overline{\mu}}(\G p;t_i,t_f)|^2\, .
\eea
Because the Pauli principle holds, the vacuum cannot carry contributions with $\boldsymbol{k}=\boldsymbol{p}$. Then we 
isolate the contribution with $\G k= \G p$
\bea
 \mathcal{P}_{e \to e e \overline{\mu}}(\G p;\De t)  &\!\!\! = &\!\!\! \!\!\!\!\!\! \sum_{\G k \neq \G p,s_2,s_3} \!\!\!|\tilde{\mathcal{A}}^{ s_2 s_3}_{e \to e e\overline{\mu}}(\G k;t_i,t_f)|^2+ \sum_{s_2,s_3} |\tilde{\mathcal{A}}^{ s_2 s_3}_{e \to e e\overline{\mu}}(\G p;t_i,t_f)|^2-\sum_{s_3} \!\!\!|\tilde{\mathcal{A}}^{r s_3}_{e \to e e\overline{\mu}}(\G p;t_i,t_f)|^2\non \\[2mm]
& = & \!\!\! \!\!\!\!\!\! \sum_{\G k \neq \G p,s_2,s_3} |\tilde{\mathcal{A}}^{ s_2 s_3}_{e \to e e\overline{\mu}}(\G k;t_i,t_f)|^2+\sum_{s_3} |\tilde{\mathcal{A}}^{r s_3}_{e \to e e\overline{\mu}}(\G p;t_i,t_f)|^2 \, .
\eea 
For large-$V$
\be
\mathcal{P}_{e \to e e \overline{\mu}}(\G p;\De t)  \ = \ V \sum_{s_2,s_3} \, \int \!\! \frac{\dr^3 \G k}{(2 \pi)^3} \, |\tilde{\mathcal{A}}^{ s_2 s_3}_{e \to e e \overline{\mu}}(\G k;t_i,t_f)|^2+\sum_{s_3} |\tilde{\mathcal{A}}^{r s_3}_{e \to e e\overline{\mu}}(\G p;t_i,t_f)|^2 \, . 
\label{thirda}
\ee
Therefore only the second piece gives a finite relevant contribution. The final result is
\be 
\mathcal{P}_{e \to e e \overline{\mu}}(\G p;\De t)  \ = \  \frac{4 m_{e \mu}^2 Y_\G p^2 }{\lf(\om_{\G p,e}+\om_{\G p,\mu}\ri)^2}  \sin^2\lf(\frac{\lf(\om_{\G p,\mu}+\om_{\G p,e}\ri)\De t}{2}\ri)  \, ,
\label{thirdb}
\ee
where
\be
Y_{\G p}^2 \ = \ \sum_{\G s} \lf(Y^{rs}_{\G p}\ri)^* Y^{rs}_{\G p} \, ,
\ee
and
\be
Y_{\G p} \ = \ \frac{|\G p|}{\sqrt{4 \om_{\G p,e}\om_{\G p,\mu}}}
\lf(\sqrt{\frac{\om_{\G p,\mu}+m_\mu}{\om_{\G p,e}+m_e}}+\sqrt{\frac{\om_{\G p,e}+m_e}{\om_{\G p,\mu}+m_\mu}}\ri) \, ,
\ee
see Fig. \ref{peeantimu} for an numerical example. Note, its effect is much smaller than the first decay process depicted in Fig/ \ref{pemu}. A direct comparison of both contributions for small time intervals can be found in Fig. \ref{comparison}.

The total decay probability of $\nu_e$ is given by the sum of Eq.\eqref{wcon} and \eqref{thirdb} \cite{Blasone:2023brf}:
\bea \non
\hspace{-1cm}\mathcal{P}^{e}_{D}(\G p; \De t) & = & 4 m^2_{e \mu} \lf[ \frac{W_{\G p}^2}{\lf(\om_{\G p,e}-\om_{\G p,\mu}\ri)^2}  \sin^2\lf(\frac{\lf(\om_{\G p,\mu}-\om_{\G p,e}\ri)\De t}{2}\ri) \ri.\;\;\;\;\;\;\;\; \\[2mm]
& & \;\;\;\;\;\;\;\;\;\;\;\; \;\;\;\;\;\;\;\;\;\;\;\;\;\;\;+  \lf. \frac{Y_\G p^2 }{\lf(\om_{\G p,e}+\om_{\G p,\mu}\ri)^2}  \sin^2\lf(\frac{\lf(\om_{\G p,\mu}+\om_{\G p,e}\ri)\De t}{2}\ri) \ri] \, . 
\eea
For short-time intervals
\bea \label{st1}
\hspace{-1cm}\mathcal{P}^{e}_{D}(\G p; \De t) & \approx & 2 m^2_{e \mu} \De t^2 \, . 
\eea
The Feynman diagrams of the above processes are shown on the l.h.s. of figure \ref{diagrams}.

We now define
\bea
 |U_\G p| & = & W_{\G p}\frac{m_\mu-m_e}{\om_{\G p,e}-\om_{\G p,\mu}} \ = \ \sqrt{\frac{\lf(\om_{\G p,e}+m_e\ri)\lf(\om_{\G p,\mu}+m_\mu\ri)}{4\om_{\G p,e}\om_{\G p,\mu}}}  \lf(1+\frac{|\G p|^2}{\lf(\om_{\G p,e}+m_e\ri)\lf(\om_{\G p,\mu}+m_\mu\ri)}\ri)\, ,\\[2mm]
|V_\G p| & = & Y_{\G p}\frac{m_\mu-m_e}{\om_{\G p,e}+\om_{\G p,\mu}} \ = \ \sqrt{\frac{\lf(\om_{\G p,e}+m_e\ri)\lf(\om_{\G p,\mu}+m_\mu\ri)}{4\om_{\G p,e}\om_{\G p,\mu}}} \lf(\frac{|\G p|}{\om_{\G p,e}+m_e}-\frac{|\G p|}{\om_{\G p,\mu}+m_\mu}\ri)\, .
\eea
Then, we can write the decay probability as
\be \label{mprob}
\mathcal{P}^{e}_{D}(\G p; \De t)  \ = \ \sin^2 2 \theta \lf[ |U_\G p|^2 \sin^2\lf(\frac{\lf(\om_{\G p,\mu}-\om_{\G p,e}\ri)\De t}{2}\ri)+ |V_\G p|^2  \sin^2\lf(\frac{\lf(\om_{\G p,\mu}+\om_{\G p,e}\ri)\De t}{2}\ri) \ri] \, . 
\ee
with $\theta=m_{e \mu}/(m_\mu-m_e) \approx \sin \theta$.
In the approximation we used, this coincides with the oscillation probability derived in the flavor Fock space approach \cite{Blasone:1995zc,BHV99,Smaldone:2021mii}.

The survival probability can be computed considering
\be
|\nu^r_{\G p,e}\ran \ \rightarrow \ |\nu^s_{\G k,e}\ran \, .
\ee
The relevant Feynman diagrams are shown on the r.h.s. of figure \ref{diagrams}.
The result is the expected one 
\be
\mathcal{P}^{e}_{S}(\G p; \De t) \ = \   1-\sin^2 2 \theta \lf[ |U_\G p|^2 \sin^2\lf(\frac{\lf(\om_{\G p,\mu}-\om_{\G p,e}\ri)\De t}{2}\ri)+ |V_\G p|^2  \sin^2\lf(\frac{\lf(\om_{\G p,\mu}+\om_{\G p,e}\ri)\De t}{2}\ri) \ri] \, , 
\ee
which preserves unitarity
\be
\mathcal{P}^{e}_{D}(\G p; \De t) + \mathcal{P}^{e}_{S}(\G p; \De t)\ = \ 1 \, .
\ee
Note that, from Eq.\eqref{st1}
\bea \label{st2}
\hspace{-1cm}\mathcal{P}^{e}_{S}(\G p; \De t) & \approx & 1-2 m^2_{e \mu} \De t^2 \, . 
\eea
Then, for short-time intervals, neutrino oscillations behave akin decays (see Eq.\eqref{pest}).
\section{Conclusions} \label{conc}

In this paper we have explored the similarities in the behavior of unstable particles and oscillating neutrinos through perturbation theory within the interaction picture of quantum field theory. We have seen how to study  decaying systems in the interaction picture and then we illustrated how analogous calculations can be used to determine the transition probabilities for flavor oscillations. Interestingly, the expressions for neutrino oscillations and particle decays are identical in the short-time range. Additionally, we demonstrated that the flavor oscillation formula derived using this method aligns, within the adopted approximation, with the one obtained via the flavor Fock space approach.

Let us emphasize the importance of focusing on the time evolution operator rather than the $S$-matrix. This approach is crucial because flavor oscillations can only be accurately described over finite time intervals. In fact, flavor neutrino states are not well-defined as asymptotically stable states. As seen in the previous results, taking the limits $t_i \to -\infty$ and $t_f \to +\infty$ eliminates flavor-changing processes while ensuring strict energy conservation. This highlights the growing significance of studying finite-time quantum field theory (QFT) \cite{Anselmi:2023wjx,Anselmi:2023phm}.
\section*{Acknowledgements}
\vspace{2mm}
G.T. would like to thank FAPESP  2023/06278-2 and CNPq bolsa de produtividade 305731/2023-8 for the financial support.
\section*{References}
\vspace{2mm}


\bibliography{LibraryNeutrino}{}

\providecommand{\newblock}{}
\begin{thebibliography}{10}
\expandafter\ifx\csname url\endcsname\relax
  \def\url#1{{\tt #1}}\fi
\expandafter\ifx\csname urlprefix\endcsname\relax\def\urlprefix{URL }\fi
\providecommand{\eprint}[2][]{\url{#2}}

\bibitem{Blasone2019}
Blasone M, Jizba P and Smaldone L 2019 {\em Phys. Rev. D\/} {\bf 99}

\bibitem{Blasone:2020qbo}
Blasone M, Jizba P and Smaldone L 2020 {\em J. Phys. Conf. Ser.\/} {\bf 1612} 012004

\bibitem{Blasone2020}
Blasone M, Lambiase G, Luciano G~G, Petruzziello L and Smaldone L 2020 {\em Class. Quant. Grav.\/} {\bf 37} 155004

\bibitem{Luciano:2023yzt}
Luciano G~G and Smaldone L 2023 {\em Symmetry\/} {\bf 15} 2032

\bibitem{Bhattacharyya_1983}
Bhattacharyya K 1983 {\em Journal of Physics A: Mathematical and General\/} {\bf 16} 2993

\bibitem{Giunti:1993se}
Giunti C, Kim C~W, Lee J~A and Lee U~W 1993 {\em Phys. Rev. D\/} {\bf 48} 4310

\bibitem{Jacob1961}
Jacob R and Sachs R~G 1961 {\em Phys. Rev.\/} {\bf 121}(1) 350--356

\bibitem{Blasone:2023brf}
Blasone M, Giacosa F, Smaldone L and Torrieri G 2023 {\em Eur. Phys. J. C\/} {\bf 83} 736

\bibitem{Blasone:1995zc}
Blasone M and Vitiello G 1995 {\em Annals Phys.\/} {\bf 244} 283 [Erratum: Annals Phys. 249, 363 (1996)]

\bibitem{BHV99}
Blasone M, Henning P~A and Vitiello G 1999 {\em Phys. Lett. B\/} {\bf 451} 140

\bibitem{Blasone:2001qa}
Blasone M, Jizba P and Vitiello G 2001 {\em Phys. Lett. B\/} {\bf 517} 471

\bibitem{Smaldone:2021mii}
Smaldone L and Vitiello G 2021 {\em Universe\/} {\bf 7} 504

\bibitem{PhysRevLett.71.2687}
Bernardini C, Maiani L and Testa M 1993 {\em Phys. Rev. Lett.\/} {\bf 71}(17) 2687

\bibitem{facchi1999regola}
Facchi P and Pascazio S 1999 {\em La regola d'oro di Fermi\/} Quaderni Di Fisica Teorica (Bibliopolis)

\bibitem{Giacosa:2010br}
Giacosa F and Pagliara G 2011 {\em Mod. Phys. Lett. A\/} {\bf 26} 2247

\bibitem{Giacosa:2011xa}
Giacosa F 2012 {\em Found. Phys.\/} {\bf 42} 1262--1299

\bibitem{Giacosa:2018dzm}
Giacosa F 2018 {\em Adv. High Energy Phys.\/} {\bf 2018} 4672051

\bibitem{Giacosa:2021hgl}
Giacosa F 2022 {\em Phys. Lett. B\/} {\bf 831} 137200

\bibitem{Giacosa:2007bn}
Giacosa F and Pagliara G 2007 {\em Phys. Rev. C\/} {\bf 76} 065204

\bibitem{Anselmi:2023wjx}
Anselmi D 2023 {\em JHEP\/} {\bf 07} 099

\bibitem{Anselmi:2023phm}
Anselmi D 2023 {\em JHEP\/} {\bf 07} 209

\end{thebibliography}
\bibliographystyle{iopart-num}
\end{document}